\documentclass[showpacs,prb,superscriptaddress,twocolumn]{revtex4}
\usepackage{textcomp}
\usepackage{amsmath}
\usepackage{amssymb}
\usepackage{graphicx}

\usepackage{color}
\usepackage{ulem}
\usepackage{textcomp}

\begin{document}

\title{Proposal for a Datta-Das transistor in the quantum Hall regime} 

\author{Luca Chirolli, Davide Venturelli, Fabio Taddei, Rosario Fazio, and Vittorio Giovannetti}

\affiliation{NEST, Scuola Normale Superiore and Istituto Nanoscienze-CNR, I-56127 Pisa, Italy}

\begin{abstract}
We propose a resonant spin-field-effect transistor for chiral spin-resolved edge states in the integer 
quantum Hall effect with Rashba spin-orbit interaction. It employs a periodic array of voltage-controlled 
top gates that locally modulate the Rashba spin-orbit interaction. Strong resonant spin-field-effect is achieved 
when the array periodicity matches the inverse of the wave-vector difference of the two chiral states 
involved. 
Well known techniques of separately contacting the edge states make possible to selectively populate and 
read-out the edge states, allowing full spin read-out. The resonant nature of the spin-field effect and the 
adiabatic character of the edge states guarantee a high degree of robustness with respect to disorder. 
Our device represents the quantum Hall version of the  all-electrical Datta-Das spin-field effect transistor.
\end{abstract}

\maketitle

\section{Introduction}

Research in spintronics~\cite{Fabian}, a spin based electronics, has led to a number of fundamental 
discoveries in physics and it has a clear potential impact in the emerging quantum technologies, such 
as quantum information processing~\cite{LossDiVincenzo98}. Among  the most intriguing proposals in 
this context  there is the spin-field-effect transistor (SFET) introduced by Datta and Das~\cite{DattaDas} 
(for a review see also Ref.~[\onlinecite{Fabian}] and references therein).  Spin-polarized carriers are selectively injected 
in a two-dimensional electron gas (2DEG), while spin precession  is externally controlled by means of a 
local gate voltage which modulates  the Rashba Spin-Orbit Interaction (SOI)~\cite{Rashba60,Rashba84}. 
The possibility of achieving gate control of the SOI strength has been experimentally established since 
the end of the '90s -- see e.g. Refs.~[\onlinecite{NittaPRL97,MillerPRL03,MeierNatPhys07}]. Spin-polarized 
injection into a semiconductor proved instead to be a more difficult task~\cite{SCHMIDT,Fabian}  which 
has  been partially overcome only recently~\cite{SPIN} delaying the first implementation of a SFET by 
almost a decade~\cite{Koo}.

The challenging requirements of the SFET can be more naturally met in a quantum Hall system. In the present 
work we introduce a generalization of the Datta-Das transistor in which gate-controlled spin procession is 
accompanied by coherent electron transfer between co-propagating chiral edge channels of the integer 
quantum Hall effect (IQHE)~\cite{DATTA}. Due to the possibility of individually contacting chiral edge 
states that have opposite values of the spin degree of freedom, our proposal completely avoids the injection 
problems~\cite{SCHMIDT,SPIN} one faces in the original SFET setting. Furthermore, our implementation of 
the SFET represents a useful tool available for quantum Hall interferometry~\cite{Heiblum2003,ROU,LIV}, 
since it can also be employed as a tunable beam-splitter, especially in the newly introduced co-propagating 
architecture of Ref.~[\onlinecite{VG}]. As such it constitutes a potentially scalable building block for implementing quantum 
information processing~\cite{STACE,IONICIOU,BERTONI} based on dual-rail encoding schemes~\cite{DUALRAIL}.
Finally, chiral states of the IQHE are ideal for long distance transfer of spin states, owing to the very large 
relaxation~\cite{MullerPRB1992,Karmakar2011} and coherence~\cite{Heiblum2003} lengths. 

\begin{figure}[t]
 \begin{center}
  \includegraphics[width=8cm]{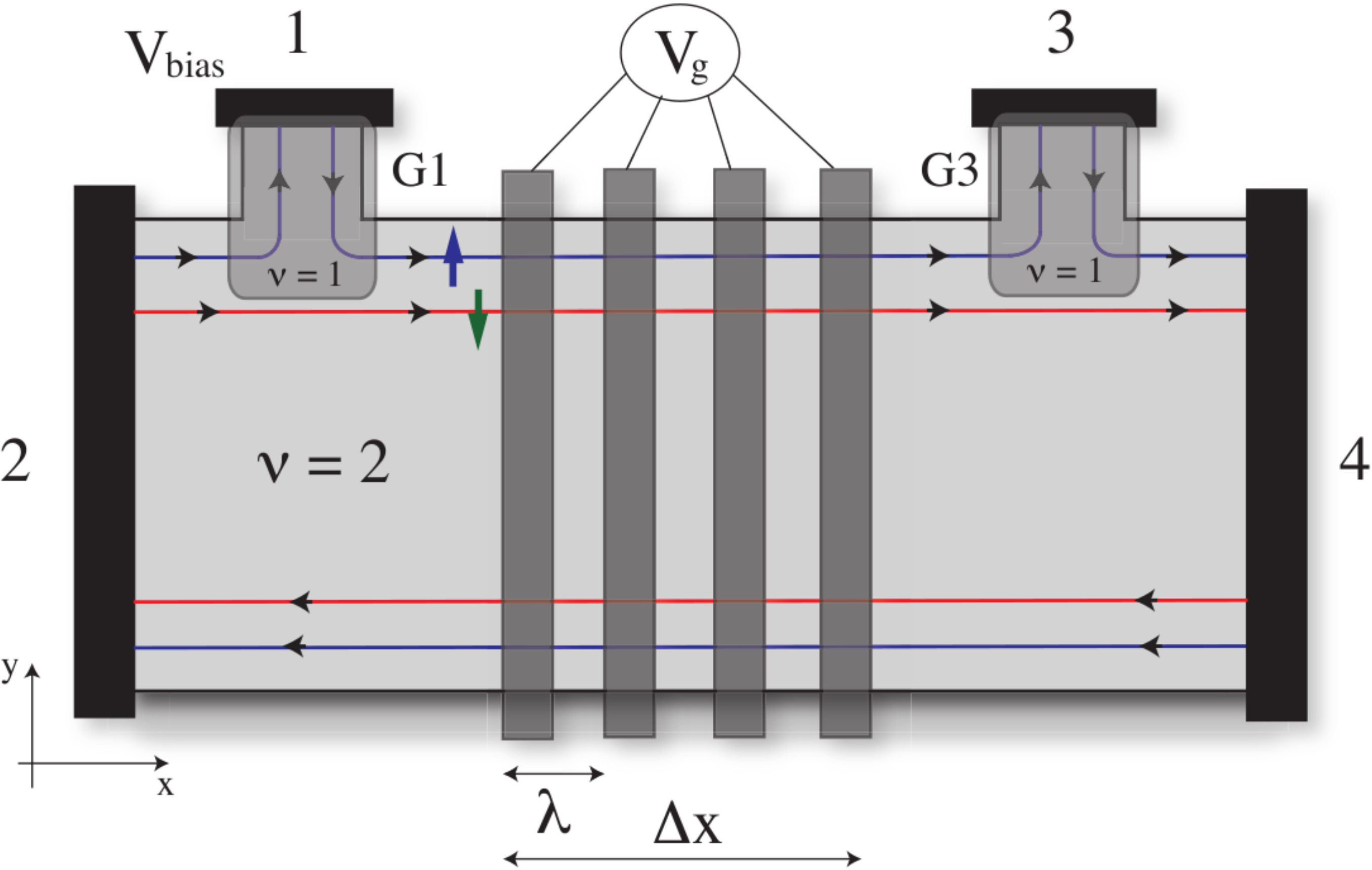}
    \caption{(Color online) Schematics of the setup: black and gray elements
    are contacts and top gates, respectively, while the chiral edge states  are represented by
  blue (outer edge $\psi_o$, spin-up) and red  (inner edge $\psi_i$, spin-down) trajectories.  
  The periodic array of top gates controlled by the voltage $V_g$ represents the scattering active region. 
    \label{fig1}}
 \end{center}
\end{figure}

The quantum Hall SFET discussed here operates  in the  IQHE regime at filling factor $\nu=2$ (number of 
filled energy levels in the bulk) in the presence of a weak uniform Rashba SOI. The two orthogonal chiral states 
$\psi_{o}$ and  $\psi_{i}$ belonging to the lowest Landau level co-propagate at slightly different wave vectors, 
$k_{o}$ and $k_{i}$, spatially separated along the transverse $y$-direction, and have (almost) opposite $z$-spin components. The underlying principle of operation can be summarized as follows (the setup  is schematized 
in Fig.~\ref{fig1}). Exploiting large top gates G1 and G3 that locally lower the filling factor to $\nu=1$, it is possible 
to separately contact $\psi_{o}$ and $\psi_{i}$. Specifically, spin-up electrons (in $\psi_{o}$) can be injected by 
applying a bias voltage $V_{\rm bias}$ to contact 1, leaving the other three contacts grounded, while a 
measurement of the charge current in contact 3 and 4 allows to determine the spin in a spin-charge conversion fashion. 
Spin precession occurs if the difference $\Delta k=k_o-k_i$ in momentum of the two chiral states is compensated 
(this is not required in the original Datta and Das setting). To tackle this problem we adopt a periodic-poling technique similar to that one experimentally realized in Ref.~[\onlinecite{Karmakar2011}] where an engineered coupling between co-propagating chiral spin-resolved edge states was achieved by means of an array of magnetic Cobalt contacts. 

Specifically, we consider a periodic array of electric top gates, placed between contacts 1 and  3 of the device 
(see Fig.~\ref{fig1}), which generate a periodic electrostatic potential and modulate the Rashba SOI 
in the 2DEG region underneath them. Under these conditions, we show that 
when the periodicity $\lambda$ of the modulation is adjusted at the resonant condition 
$\lambda_{\rm res}=2\pi /\Delta k$ a strong enhancement of the spin transfer is obtained and it can be 
externally controlled by varying the top gate voltage (typical values of  $\lambda_{\rm res}$ are hundreds 
of nanometers in GaAs, easily experimentally achievable -- see e.g. Refs.~[\onlinecite{Karmakar2011,KouwenhovenPRL90}]).
As a result, in the same spirit of the original Datta and Das scheme~\cite{DattaDas}, our method provides an 
{\it electrically tunable}  way to achieve a resonant spin-field-effect transistor in quantum Hall systems.

The article is structured as follows. In Sec.~\ref{Model} we introduce the model and define its Hamiltonian. In Sec.~\ref{Transport} we address the transport properties of the system. In Sec.~\ref{Disorder} we test the robustness of the device in the presence of disorder. Finally we conclude in Sec.~\ref{Conclusions}.

\section{The model}
\label{Model}

A rigorous analysis of the effect is obtained by expressing the chiral edge states $\psi_{o}$ and $\psi_{i}$  as  eigenvectors
of the unperturbed Hamiltonian $H = H_0 + H_R$, where 
\begin{equation}
H_0 =\frac{\boldsymbol{\Pi}^2}{2m^*}+V_c(y)-\frac{E_z}{2}\sigma_z\;, 
\end{equation}
describes a 2DEG in the presence of a strong perpendicular magnetic field $B$ and of a  transverse 
confining potential $V_c(y)$, and 
\begin{equation}
H_R=\alpha(\Pi_x\sigma_y-\Pi_y\sigma_x)\;,
\end{equation}
is the weak uniform Rashba SOI contribution. The Rashba coupling constant is denoted by $\alpha$, 
while $\sigma_{x,y,z}$ are the spin Pauli matrices. ${\bf \Pi}=-i\hbar\boldsymbol{\nabla}+e{\bf A}$ is the 2D 
kinetic momentum in the Landau gauge ${\bf A}=(-By,0,0)$,  $E_z=g^*\mu_BB$ the Zeeman gap, $m^*$ 
and $g^*$ the effective mass and $g$-factor of the material, respectively. The effect of a uniform Rashba 
SOI in the IQHE has been intensely  studied in the past~\cite{Winkler,PalaGovernale2005,Grigoryan09,REYNO}.
The eigenstates $\psi_{o}$ and $\psi_{i}$ are given in terms of dressed states composed by linear 
combination of adjacent Landau levels of opposite spin \cite{Winkler}. Defining the cyclotron gap 
$E_c=\hbar\omega_c$, with  $\omega_c=eB/m^*$ the  cyclotron frequency, and the magnetic length 
$\ell_B=\sqrt{\hbar/eB}$, at first order in $\hbar\alpha/\ell_BE_c$ the edge states are 
\begin{eqnarray}
\psi_{o} &\approx& \psi_{0,k_o,\uparrow}+i \frac{\sqrt{2} \hbar\alpha}{\ell_B(E_c+E_z)}\psi_{1,k_o,\downarrow},\label{Eq:psiOuter}\\
\psi_{i} & = & \psi_{0,k_i,\downarrow},\label{Eq:psiInner}
\end{eqnarray}
where $\psi_{n,k,\sigma}$ are  the eigenvectors of $H_0$ associated with the $n$-th Landau level, longitudinal momentum $k$, and having spin $z$-projection equal  to $\sigma$. 
Under the assumption of  smooth confinement $V_c(y)$, the latter can be written as 
$\psi_{n,k,\sigma}(x,y)\simeq e^{ikx} \chi_n(y-k\ell_B^2)|\sigma\rangle/\sqrt{L}$, with $\chi_n(y)$ being 
the $n$-th eigenfunction of the harmonic oscillator  and 
$L$ is the length of the Hall bar in the $x$-direction. In a single-particle picture, the values  $k_o$ and $k_i$ can be estimated from the degeneracy condition 
$V_c(k_{o,i}\ell_B^2)-m^*\alpha^2(1\pm 1)/(1+E_z/E_c)\mp E_z/2=E_F-E_c/2$, where $E_F$ 
is the Fermi energy of the system, the ratio $E_z/E_c=m^*g^*\mu_B/e\hbar$ being a 
material-dependent constant. The momenta difference is  
$\Delta k=k_{o}-k_{i}\simeq  ( E_z+m^* \alpha^2/(1+E_z/E_c) )/(\ell_B^2 \partial_yV_c)|_{E_F}$, 
and  the  transverse spatial separation  of the edge states is  $\Delta y=\Delta k \ell_B^2$. The latter is substantially renormalized by exchange interaction  \cite{DempseyHalperin}, which we effectively take into account in the renormalization of the $g$-factor.  According to this analysis  the $\psi_{o}$ edge state will carry 
essentially spin-up electrons at momentum $k_o$, while $\psi_{i}$ purely spin-down electrons 
at momentum $k_i$. 

\subsection{Modulated Rashba SOI}
\label{PeriodicPert}

The perturbation  introduced by the top gate array  of  Fig.~\ref{fig1} (which extends in the longitudinal 
$x$-direction for a region of length $\Delta x \ll L$) produces two net effects, both contributing to electron 
transfer  from $\psi_o$ to $\psi_i$. First of all, the array of top gates induces a periodic modulation of the 
electrostatic potential underneath the gates that can be described by adding to the Hamiltonian a term 
\begin{equation}
U_m(x)=U(V_g)\cos^2(\pi x/\lambda). 
\end{equation}
The latter locally mixes the Landau levels and it can open 
mini-gaps in the spectrum. However, as long as the potential $U_m$ is restricted to a small region 
$\Delta x$ and  $U(V_g) \ll E_c$, no qualitative change is expected in the structure of edge 
states~\cite{footnote}. In this regime, even though $U_m$ does not couple to the spin degree 
of freedom of the system, it can  contribute indirectly to the coupling between $\psi_o$ and $\psi_i$ 
due to the modification of the edge states induced by  the homogeneous  $H_R$ term in the Hamiltonian. 
Furthermore, a  second important effect introduced by the gate array  is a periodic modulation of the 
Rashba coupling constant~$\alpha$. We  describe it  as an additional perturbation term to $H$ of the form 
\begin{equation}\label{Eq:Rashba-x}
\delta H_R(x)=\frac{1}{2}\{\delta\alpha(x),\Pi_x\}\sigma_y-\delta\alpha(x)\Pi_y\sigma_x\;,
\end{equation}
where  the modulation $\delta\alpha(x)=\delta\alpha(V_g)\cos^2(\pi x/\lambda)$ has the same spatial
dependence as $U_m(x)$ but strength $\delta\alpha(V_g)$, and where the symmetrized product 
$\{\delta\alpha(x),\Pi_x\}/2=(\delta\alpha(x)\Pi_x+\Pi_x\delta\alpha(x))/2$ is inserted to ensure 
hermiticity of the SOI term. Similar contributions  have been analyzed for spin transport in quantum 
wires, both in a single particle-picture \cite{Wang2004} and for an interacting nanowires 
\cite{Japaridze2009} and helical modes in the quantum spin Hall effect \cite{Strom2010}. 
The  $\delta H_R(x)$ term provides a direct spin-flip mechanism that is resonantly enhanced by 
the periodicity of the gate array and couples $\psi_{o}$ and~$\psi_{i}$. According to 
Refs.~[\onlinecite{NittaPRL97,Koo}], $\alpha$ is approximatively a linear function of $V_g$ in InAs-based 
heterostructures. More precisely, for a large top gate $\hbar\alpha$ varies within 
$[0.7,0.9]~\times~10^{-11}$ eVm in the voltage range $V_g=[0,-1]$ V~\cite{NittaPRL97}, and 
$\hbar\alpha\sim [0.9,1.3]~\times~10^{-11}$ eVm in the voltage range $V_g=[0,-3]$ V~\cite{Koo}. 
It follows that $\delta \alpha(V_g)$ can be tuned almost by a factor 2 and that 
$\delta\alpha(V_g)/\alpha \lesssim 1$ for all practical purposes.

\section{Transport}
\label{Transport}

We characterize the transport properties of the system in the Landauer-B\"uttiker formalism 
\cite{LandauerBuettiker}.  The current flowing through the output contacts can be 
determined knowing the amplitudes for scattering off the region underneath 
the top gate array  of Fig.~\ref{fig1}. Asymptotic states are defined far away from the scattering region and  are given by the edge states of the Hall 
bar $\psi_o$ and $\psi_i$, Eqs.~(\ref{Eq:psiOuter},\ref{Eq:psiInner}). 
The scattering amplitudes, at the Fermi energy $E_F$, can be generally expressed 
in terms of the matrix elements of the operator 
$T=\Delta H+\Delta H G^{+}(E_F)\Delta H$,~\cite{DiVentra}  
where $G^{+}(\epsilon)=(\epsilon+i0^+-H-\Delta H)^{-1}$ is the full retarded 
Green's function of the system, $\Delta H= U_m(x) + \delta H_R(x)$ is the scattering potential (finite 
only in a region of length $\Delta x$) and $H+\Delta H$ is the overall 
Hamiltonian of the system. In particular, the transmission amplitude for 
scattering from the incoming state $\psi_i$ to the outgoing states $\psi_o$ is given by
$t_{io}=(L/i\hbar\sqrt{|v_i||v_o|})\langle\psi_o|T|\psi_i\rangle$, 
where $v_i$ and $v_o$ are the velocities of the incoming and outgoing states, respectively.

Let us assume first that the term $\Delta H$ can be treated as a perturbation with respect to $H$ 
(i.e. an assumption valid for small $V_g$). Setting $v_i=v_o\equiv v_D=\partial V_c(k\ell_B^2)/\partial \hbar k|_{E_F}$ 
the drift velocity of the system, we can evaluate the scattering amplitude $t_{io}$ in the Born approximation
$t_{io}\simeq (L/i\hbar v_D)\langle\psi_{i}|\Delta H |\psi_{o}\rangle$ and we find
\begin{equation}\label{Eq:BornApp}
t_{io}=t(V_g)~{\rm sinc}((\Delta k-2\pi/\lambda)\Delta x/2)\;,
\end{equation}
where $\text{sinc}[\cdot]=\sin[\cdot]/[\cdot]$ is the cardinal sinus, and where 
\begin{equation}\label{T12SW}
t(V_g)=\frac{\Delta x \Delta k}{2v_D}\left[i\delta\alpha(V_g)-\frac{U(V_g)}{E_c+E_z}
\alpha\right]e^{-\frac{\Delta k^2\ell_B^2}{4}}\;.
\end{equation}
This is the main result of the paper. 
Equation~(\ref{Eq:BornApp}) clearly shows how the periodicity of the gate array can provide the 
correct amount of momentum transfer giving rise to resonant coupling of the edge states at 
$\lambda = \lambda_{\rm res}$. Equation~(\ref{T12SW}) instead confirms that both $U_m(x)$ 
and $\delta H_R(x)$ contribute to the coupling and shows that, as in the original Datta-Das 
spin-field transistor, the intensity of the coupling can be tuned electrically with the gate voltage 
$V_g$ [both $U(V_g)$ and $\delta \alpha(V_g)$~\cite{NittaPRL97,Koo} can be taken as linear 
function of $V_g$]. Interestingly, the two contribution in Eq.~(\ref{T12SW}) are $\pi/2$ phase-shifted 
and they cannot cancel. This can be understood noticing that the origin of the interplay between 
the electrostatic modulation potential and the uniform Rashba SOI arises from their non-zero 
commutator, $[H_R,U_m]\propto \partial_xU_m\sigma_y$, that turns out to be 
oscillating and $\pi/2$ phase-shifted with respect to $U_m$ and to $\delta H_R(x)$.

\begin{figure}[t]
 \begin{center}
  \includegraphics[width=9cm]{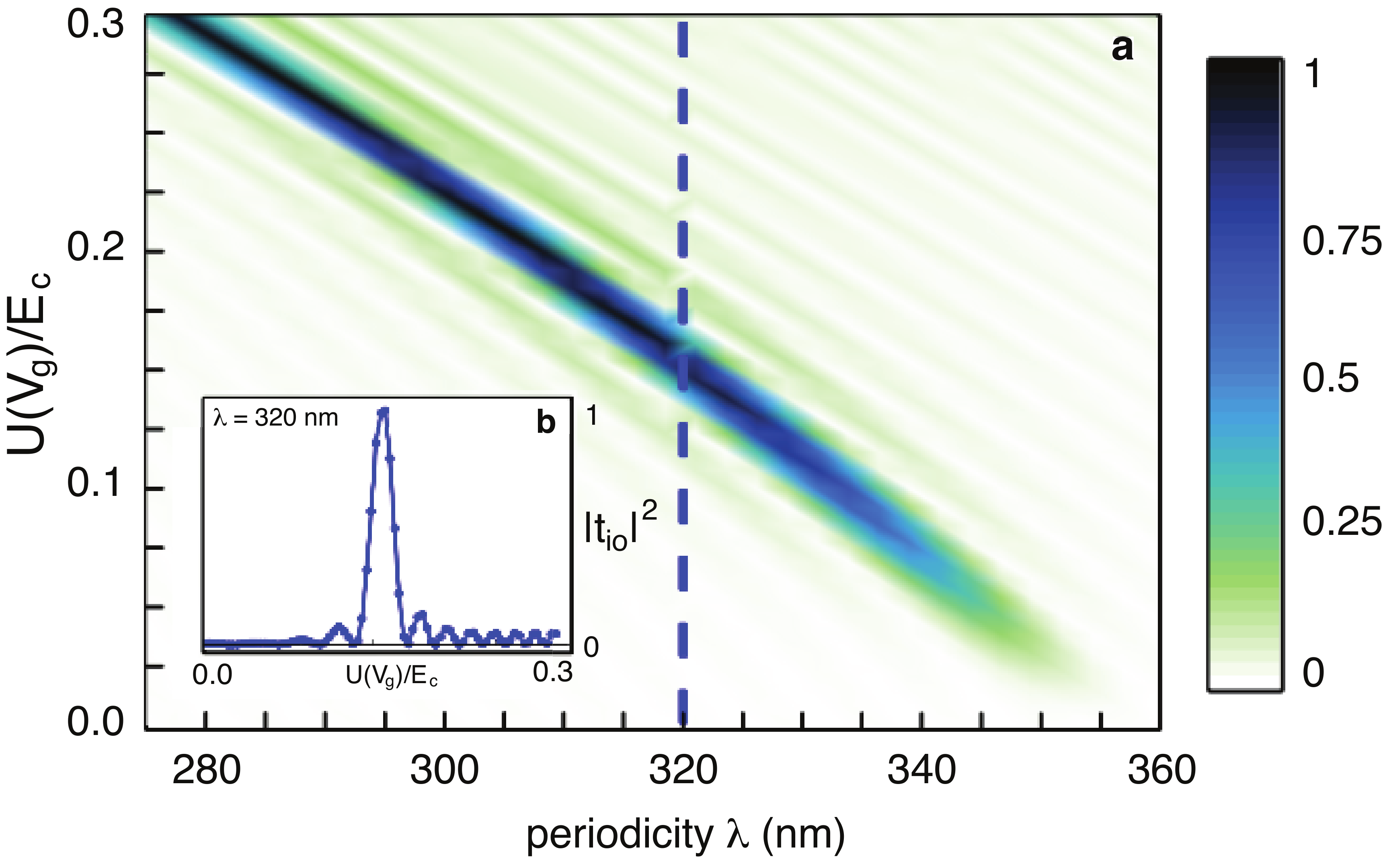}
    \caption{ (Color online) a) Density plot of $|t_{io}|^2$ versus the array periodicity $\lambda$ and the  amplitude $U(V_g)$ of the modulation potential. Both uniform and modulated Rashba SOI, ($\alpha$ and $\delta\alpha(V_g)$) are present, as detailed in the text. The total number of gates is 50 independently from $\lambda$: the abscissa spans a modulation length $\Delta x$ that ranges from 12.5 $\mu$m to 18 $\mu$m. b) $|t_{io}|^2$ corresponding to $\lambda=320~{\rm nm}$ (dashed line in the density plot). These results are obtained by means of the scattering formalism applied to the system shown in Fig.~\ref{fig1}, described on a lattice by means of the KNIT open source package~\cite{KNIT}. \label{fig}}
 \end{center}
\end{figure}

The results  obtained above are valid in the perturbative regime of weak Rashba SOI 
($\hbar\alpha/\ell_BE_c=\ell_B/\ell_{\rm so}\ll 1$, where the spin-orbit length 
$\ell_{\rm so}=\hbar/m^*\alpha$ characterizes the scale of the SOI) and weak Zeeman energy 
($E_z/E_c\ll 1$). For GaAs-based heterostructures such conditions are widely fulfilled. For InAs 
the spin-orbit length can be estimated to be of the order of hundreds on nanometers 
(using $m^*=0.02~m_0$) and, for $B\sim 5~{\rm T}$, one has $\ell_B/\ell_{\rm so}\sim 0.1$. 
The Zeeman energy is also relatively large due to the large $g$-factor, 
$g^*_{\rm InAs}\sim 3-10$ so that $E_z/E_c\sim 0.1$. Therefore, a perturbative 
analysis is relevant.

\subsection{Numerical Analysis}
\label{Numerics}

\begin{figure}[t]
\begin{centering}
\includegraphics[width=\columnwidth]{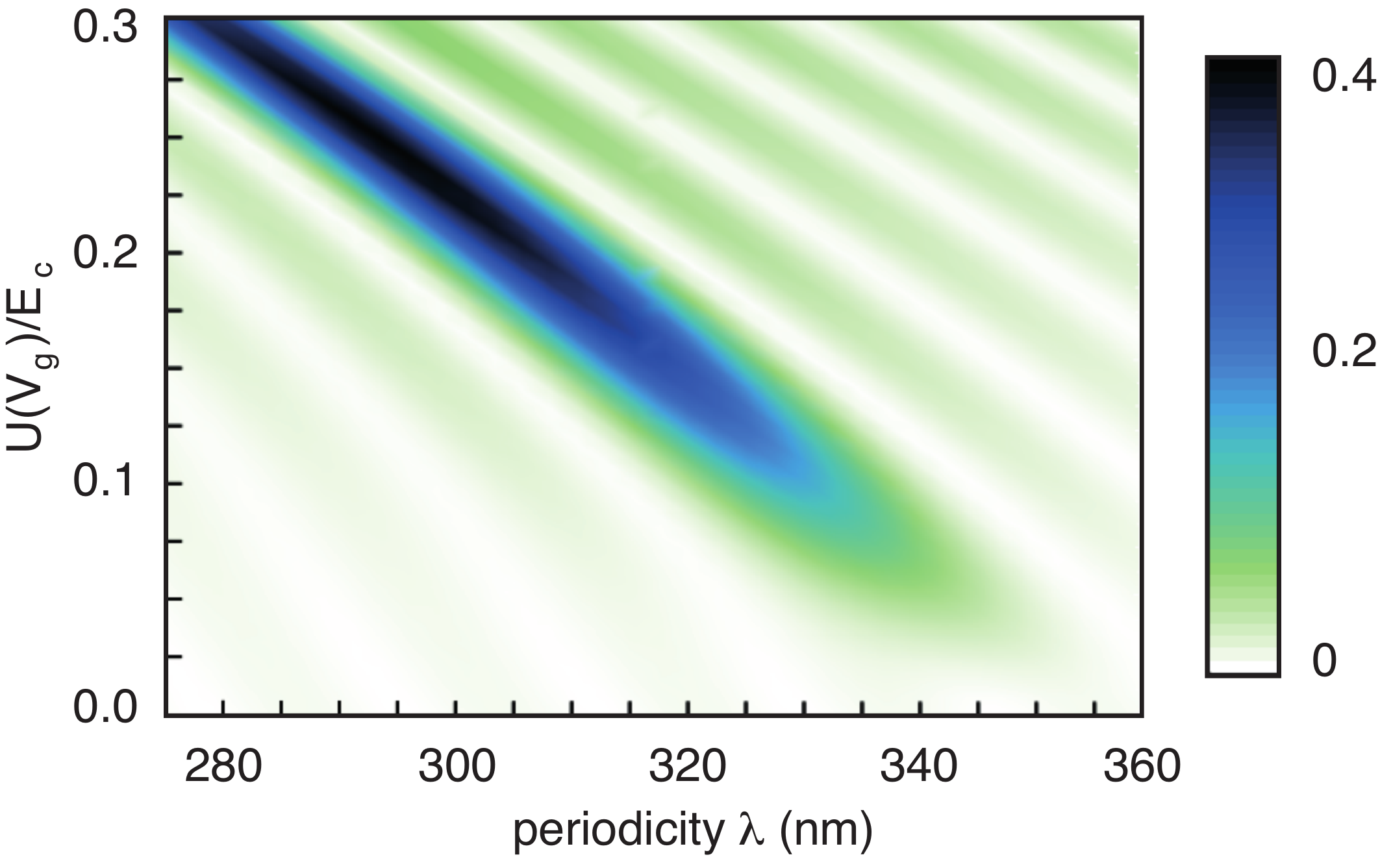}
\caption{(color online) Density plot of the transmission probability $|t_{io}|^2$ versus applied 
modulation potential $U(V_g)$ and array periodicity $\lambda$, for an array of 
20 top gates. A maximum transmission probability of 40\% is achieved for a 
periodicity $\lambda\sim 290~{\rm nm}$ for suitable applied voltage producing 
on the 2DEG a potential $U(V_g)\sim 0.25~E_c$.\label{fig3}}
\end{centering}
\end{figure}

We have checked the validity of the analytical perturbative approach by numerical calculations based on 
a tight-binding Hamiltonian describing the system in Fig.~\ref{fig1}. Transport properties are computed through the recursive-Green's function algorithm of the KNIT Numerical Package~\cite{KNIT}, which allows to compute local and non-local observables on tight-binding Hamiltonians~\cite{Ferry,DATTA}.
The total transmission between lead 1 and lead 4 of the system (see Fig.~\ref{fig1}) can be obtained from the retarded Green's functions through the 
Fisher-Lee relations~\cite{FisherLee}. The leads are assumed with no SOI. The width of the Hall 
bar is taken to be 100 nm (hard wall confining potential), 
discretized on a square lattice of spacing $2.5$ nm. The external magnetic field is taken to be $6~{\rm T}$,  corresponding to a cyclotron gap 
$\hbar\omega_c\approx 34.7~{\rm meV}$ for InAs, and it is introduced by means of Peierls 
phase factor. The Zeeman gap has been 
taken to be $E_{\rm Z}=\hbar\omega_c/15$, corresponding to an effective 
value of  $g^* = 6.6$. This value is of the order of magnitude of the 
exchange-enhanced value measured in InAs devices in the quantum 
Hall regime \cite{Desrat2004}. 

Fig.~\ref{fig}a shows the transmission probability $|t_{io}|^2$, 
in a density plot, as a function of the periodicity $\lambda$ and the amplitude $U(V_g)$ of the potential, 
for a total fixed number of top gates equal to 50. In the simulations we set 
$\hbar\alpha= 3\times10^{-12} {\rm eVm}$, and simultaneously vary $\delta\alpha(V_g)$ in the range 
$[0,\alpha]$, and $U(V_g)$ in the range $[0,\hbar \omega_{c}/3]$. The Fermi energy is taken in the middle 
of the $\nu=2$ plateau ($E_F=E_c)$, so that even for the maximum value of $U(V_g)$ backscattering is absent. 
For these parameters, in the absence of the array we obtain an intrinsic channel mixing of less than 0.1\% 
(due to the broken translation invariance owing to the geometry of the injection/detection regions in the device), 
which confirms that the effect of SOI, as such, is negligible. 

The plot shows a full resonant line in the $\lambda-U(V_g)$ plane in which spin flip is 
achieved. For moderate value of $U(V_g)$ the transmission probability reaches unity and complete spin-flip 
accompanied by total charge transfer is atteined. Beside the resonance line, a series of secondary peaks 
appears, which can be understood as interference fringes in the framework of the perturbative analysis, Eq.~(\ref{Eq:BornApp}). 
For increasing values of  the electrostatic potential $U(V_g)$ we see a shift of the resonant periodicity 
$\lambda_{\rm res}$ to smaller values. This is consistent with a local non-linear modification of the confining 
potential and  a local decrease of the Fermi energy, together yielding an increase  of the wave vector difference, 
which acquires an effective gate-voltage dependence, $\Delta k(V_g)$, entailing a ``red-shift" of the resonant 
peak $\lambda_{\rm res}=2\pi/\Delta k$. Since the value of the resonant periodicity is {\it a priori}  unknown 
and sample-dependent, such effect turns out to be a useful control knob of the setup (the range of periodicities 
at which a given sample can resonate spanning more than 100 nm). As shown in Fig.~\ref{fig}b for a fixed 
periodicity, by varying the voltage applied to the gates it is possible to control the precession of the spin, 
characteristic of a SFET. We also note that the external magnetic field represents an additional tuning 
parameter that is expected to affect the resonance condition.

For the sake of completeness we present in Fig.~\ref{fig3} results for an array of 20 top gates. The overall response presents no qualitative changes, but a smaller maximum transmission probability and interference fringes with larger periodicity, consistent with the reduced number of top gates. We 
point out that the characteristics of a transistor is a high/low current response with 
respect to a threshold. Therefore, also in the case of an array with 20 top gates 
shown in Fig.~\ref{fig3} a good transistor functionality can be achieved.

\subsection{Dresselhaus SOI}
\label{Dresselhaus}

It is worth noticing that in III-V compounds the Dresselhaus SOI, originating from 
a bulk inversion asymmetry in zinc-blende structures \cite{Dresselhaus55}, is also present. When restricted 
to quantum wells, by aligning our reference frame to the cubic crystallographic axes such that $x$, 
$y$, and $z$ correspond to the $[\bar{1}00]$, $[0\bar{1}0]$ and $[001]$, respectively, (with [001] the heterojunction growth 
direction) and neglecting cubic corrections in the momentum, the Dresselhaus Hamiltonian reads 
\begin{equation}
H_D=\beta(\Pi_x\sigma_x-\Pi_y\sigma_y),\nonumber
\end{equation}
with coupling constant $\hbar\beta\simeq 27(\pi/d)^2~{\rm eV\AA}^3$ for both GaAs and InAs \cite{Winkler}, $d$ being the width of the quantum well. In InAs, in which the fundamental gap is smaller than 
GaAs, the Rashba term dominates, whereas in GaAs, Dresselhaus and 
Rashba SOIs have comparable strength and both contribute to the modification of the edge states. In the perturbative regime 
$\alpha,\beta\ll \omega_c\ell_B$  this can be easily taken into account:  at first order in $H_R+H_D$ 
the edge states are 
\begin{eqnarray}
\psi_{o} &\approx& \psi_{0,k_o,\uparrow}+i \frac{\sqrt{2} \hbar\alpha}{\ell_B(E_c+E_z)}\;
\psi_{1,k_o,\downarrow},\nonumber\\
\psi_{i} &\approx& \psi_{0,k_i,\downarrow}+\frac{\sqrt{2}\hbar\beta}{\ell_B(E_c-E_z)}\;\psi_{1,k_i,\uparrow},\nonumber 
\end{eqnarray}
with the renormalized momenta obtained by solving 
$V_c(k_{o,i}\ell_B^2)-m^*\alpha^2(1\pm 1)/(1+E_z/E_c)-m^*\beta^2(1\mp 1)/(1-E_z/E_c)\mp E_z/2=E_F-E_c/2$, 
and the bulk gap is still dominated by the Zeeman energy. As a result the multiplicative factor in the square 
bracket Eq.~(\ref{T12SW}) acquires an extra contribution $iU(V_g)\beta/(E_c-E_z)$ which adds in phase 
to the one controlled by $\delta\alpha(V_g)$ with positive sign providing a net further increase of the 
scattering amplitude.

\section{Robustness to Disorder}
\label{Disorder}

One of the major problems in the original Datta-Das transistor is the limitation to ballistic 2DEGs. Due to the spin-orbit interaction, the spin orientation is momentum-dependent: elastic scattering randomizes the spin already after few scattering events spoiling the spin precession dynamics.
A great advantage of our proposal is the overcoming of this problem.
Indeed, the resonant nature of the spin-flip mechanism and the chiral, one-dimensional, character of  integer QH edge states suggest that the resonant SFET is robust against disorder within the adiabatic transport regime~\cite{Nonadiabatic}.
To corroborate this intuitive argument we have run numerical tight-binding simulations which include: i) the presence of imperfections in the periodicity of the fingers, ii)  the presence of corrugations at the sample edge. The results show deviations from the clean situation only for unrealistically strong disorder.

We now comment about two possible material systems (GaAs and InAs-based heterostructures), 
which suite best the implementation of our setup. Due to the larger Rashba coupling constant, a stronger 
effect could be seen in InAs-based devices. However, the larger Rashba coupling constant is also 
responsible for more sensitivity to disorder, resulting in a shorter spin-relaxation length $\ell_{\rm sr}$ 
that would limit the scalability of the device  (the ratio being estimated as 
$\ell_{\rm sr}^{\rm InAs}/\ell_{\rm sr}^{\rm GaAs}\sim 0.1$ 
\cite{KhaetskiiPRB92,PolyakovPRB96,PalaGovernale2005}). Besides, in InAs the $g$-factor can be about 
an order of magnitude larger than in GaAs. This implies a resonant periodicity of the array an order of 
magnitude smaller than the case of GaAs, which could represent a technical problem for very large $g$-factors. 

\subsection{Imperfect periodic potential}
\label{DisorderA}

\begin{figure}
\begin{centering}
\includegraphics[width=\columnwidth]{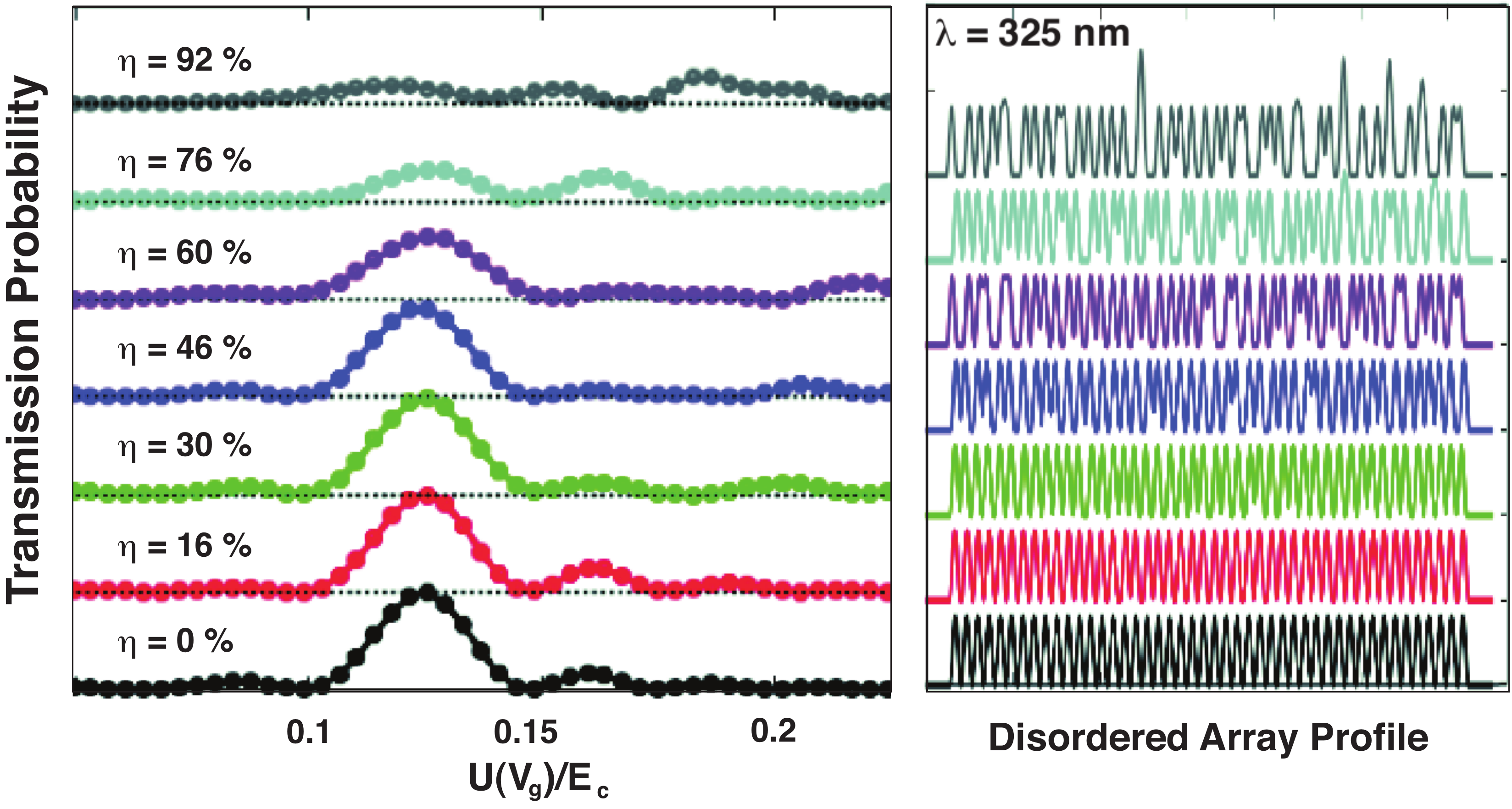}
\caption{(color online) Left Panel: plot of the transmission probability $|t_{io}|^2$ 
versus applied gate potential $U(V_g)$ for a periodicity $\lambda=325~{\rm nm}$, 
and different realizations of an imperfect periodic potential characterized by a 
degree of imperfection $\eta$. Each curve is relative to a different value of $\eta$ 
and is vertically shifted by 1: the maximum transmission is nearly unity up to  
$\eta< 46 \%$. Right Panel: Profiles of the different realizations of imperfect 
periodic potential. All non-specified parameters are equal to the ones 
used for the simulations presented in Fig.~\ref{fig1}.\label{disorderFIG1}}
\end{centering}
\end{figure}

As a first kind of disorder we consider the case of non-perfect periodicity of the 
gate array potential. This class of disorder well describes imprecisions in the 
nanofabrication of the gates as well as non-uniformities in the different sections 
of the Hall bar. We model this disorder by randomizing each finger position 
in a range $\eta \lambda$ around its ideal position. The parameter $\eta$ can thus be considered as the 
``degree of imperfection'': if $\eta=0\%$ the array is ideal, whereas if $\eta\simeq 100\%$ 
the position of each potential peak in the region  is practically random, and the 
original periodicity is severely spoiled. In Fig.~\ref{disorderFIG1} (left panel) we 
present the curves of the transmission probability versus the applied top gate 
potential $U(V_g)$  for a device with 50 top gates and periodicity 
$\lambda=325~ {\rm nm}$, for different random realizations (shown 
in the right panel of Fig.~\ref{disorderFIG1}). For degree of imperfection up to 
$\eta\sim 46\%$ we see no difference in the position and height of the main 
resonance peak, although we record an attenuation of the secondary peaks 
as we increase the degree of imperfection.  As we further increase the value of $\eta$ 
we see that the height of the main resonance peak gets lower and additional 
secondary peaks start to show up. This is due to the fact that the randomization 
procedure generates higher harmonics in the Fourier spectrum of the periodic 
potential, with reduced amplitude. The device response then corresponds to a 
sum of  weak resonances at each harmonic. The results unequivocally show 
that the resonant SFET effect survives up to strong imperfections on the top 
gates positions.

\subsection{Disordered edges}
\label{DisorderB}

\begin{figure}
\begin{centering}
\includegraphics[width=\columnwidth]{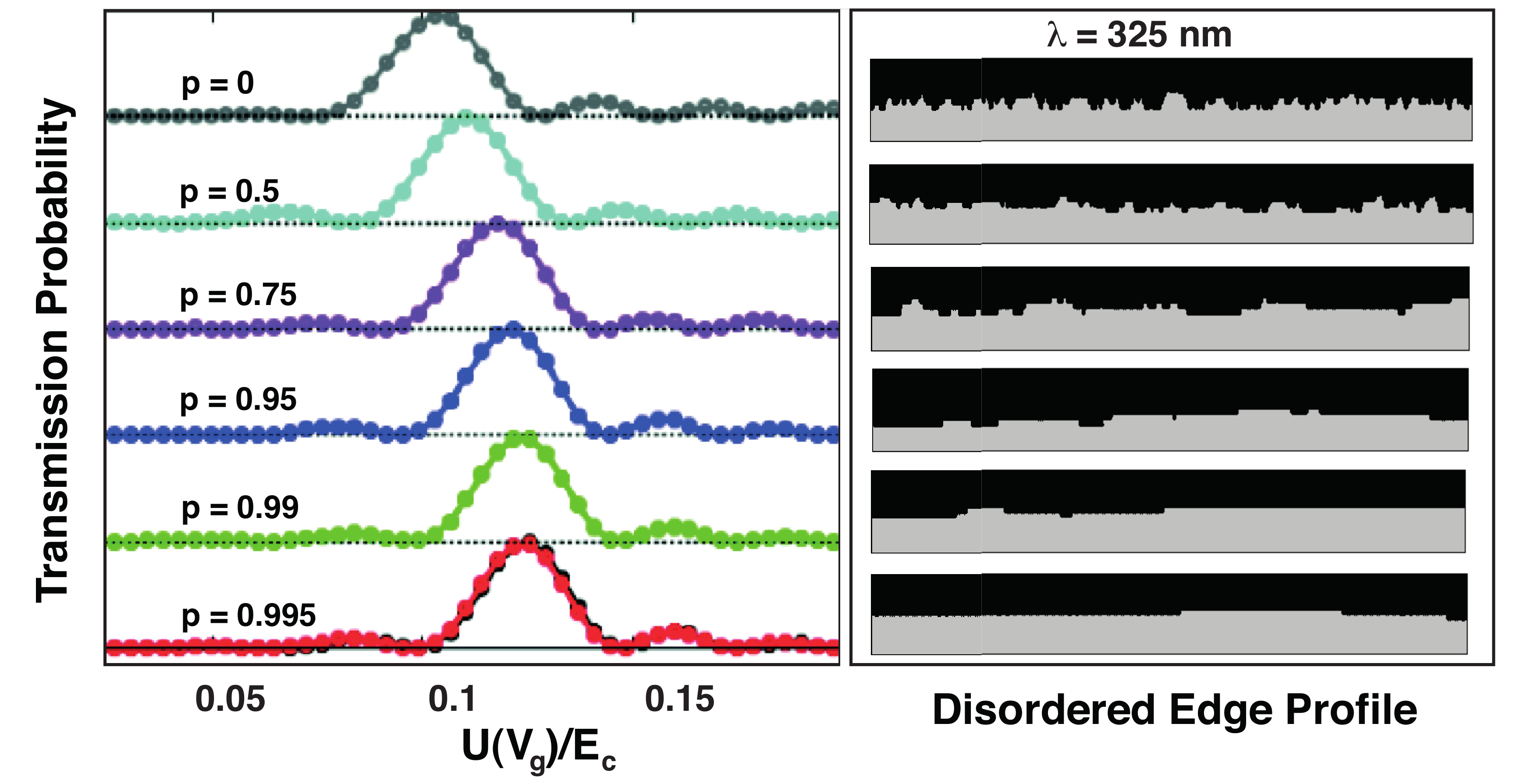}
\caption{(color online) Left Panel: plot of the transmission probability 
$|t_{io}|^2$ versus applied gate potential $U(V_g)$ for a periodicity 
$\lambda=325~{\rm nm}$, for different realizations of a rough confining 
potential characterized by the parameter $p$. Each curve is relative to a 
different value of $p$ and is vertically shifted by 1: for all the cases the 
maximum transmission reaches unity. A black curve for the case $p=1$, 
corresponding to the case of no disorder in the edge profile, is 
superimposed to the one for $p=0.995$. Right Panel: Profiles of the 
different realizations of rough confining potential. All 
non-specified parameters are equal to the ones used for the 
simulations presented in Fig.~\ref{fig1}.\label{disorderFIG2}}
\end{centering}
\end{figure}

We now consider the possibility that the edges of the sample are ``rough'', 
{\it i.e.} not perfectly straight, and we study how such a roughness affects 
the functionality of the resonant SFET proposed. Rough edges are 
implemented through a random ``corrugation'' function (see Fig.~\ref{disorderFIG2}) 
constructed with a random-walk procedure following a simple probability law. 
The latter is controlled by the parameter $p\in [0,1]$, which determines 
the characteristic length of the corrugation: for $p=1$ the profile is ideal, whereas for 
smaller $p$ it becomes progressively more corrugated. For $p=0$ 
the profile has a roughness on the scale of the lattice constant.
This is illustrated in the right panel of Fig.~\ref{disorderFIG2}, where six 
possible realizations of corrugated confining potentials, relative to as 
many values of $p$, are shown. Note that, for all the realizations considered, 
the transmission in the absence of a periodic array is negligible ($<0.05\%$).
In the left panel of Fig.~\ref{disorderFIG2} we show the corresponding 
response of the device. For all disordered potential profiles considered 
we obtain a well defined resonant transmission, with a main resonance 
peak that reaches unity for suitable gate voltage $V_g$, and a series of 
lateral secondary peaks. We note, however, a progressive shift  of the 
resonance peak to lower values of the gate potential $U(V_g)$, in 
correlation with an increase of the roughness of the edge profile and 
a consequent increase of the curvilinear abscissa of the edge channel between 
pairs of top gates. Such a shift of the resonance can be understood 
as an effective increase of the potential periodicity $\lambda$ that 
leads to a right-shift of the vertical cut in Fig.~\ref{fig} of the main text. 
We can conclude that disorder related to corrugated edges does 
not spoil the resonant character of the device, but at most moves 
the resonance.

\section{Conclusions}
\label{Conclusions}

In this paper, a resonant SFET between chiral spin-resolved edge states in the 
integer quantum Hall regime has been proposed. Coupling of the two spin states is obtained by applying an array of 
voltage-controlled top gates on the device. We notice that an analogous top-gate technique has been recently shown to represent a valuable tool for measuring the strength of the Rashba SOI in a quantum wire with uniform Rashba SOI and no-magnetic field~\cite{Thorgilsson2011}. In our case the coupling mechanism arises as a resonant  
interplay between uniform and periodically modulated Rashba SOI that shows strong robustness against 
disorder in the system.  

We acknowledge very useful conversations with F. Ciccarello, M. Governale, B. Karmakar, G. La Rocca, V. Pellegrini,  
G. Usaj, and G. Vignale. This work was supported by MIUR through FIRB-IDEAS Project No. RBID08B3FM 
and EU-Projects IP-SOLID, STREP-GEOMDISS, and NANOCTM.

\end{document}